\documentclass[12pt,twoside]{article}

\usepackage{epsfig}

\begin{document}\hbadness=10000\thispagestyle{empty}
\pagestyle{myheadings}
\markboth{H.-T. Elze}
{Emergent discrete time and quantization: relativistic particle 
with extradimensions}
\title{{\bf Emergent discrete time and quantization: relativistic particle 
with extradimensions}}
\author{$\ $\\
{\bf Hans-Thomas Elze} \footnote{Permanent address:  
Instituto de F\'{\i}sica, Universidade Federal do Rio de Janeiro, 
C.P. 68.528, 21941-972 Rio de Janeiro, RJ, Brazil }
\\ $\ $\\
Institut f\"ur Theoretische Physik, Universit\"at Heidelberg \\ Philosophenweg 16,  
D-69120 Heidelberg, Germany }
\vskip 0.5cm
\date{January 2003}
\maketitle
\vspace{-8.5cm}
\vspace*{8.0cm}
\begin{abstract}{\noindent
We study the reparametrization invariant system of a classical 
relativistic particle moving in (5+1) dimensions, of which two  
internal ones are compactified to form a torus. A discrete 
physical time is constructed based on a quasi-local invariant 
observable. Due to ergodicity, it is simply related to the proper 
time on average. The external motion in Minkowski space can then 
be described as a unitary quantum mechanical evolution.
}\end{abstract}
\section{Introduction}
Recently we have shown that for a nonrelativistic particle with time-reparametrization 
invariant dynamics one can define quasi-local observables which characterize the evolution in 
a gauge invariant way \cite{ES02}. 
  
We presently extend our study to the case of a relativistic particle, characterizing the evolution 
by statistical properties of an internal ``clock'' motion. 
Thus, ``time is change'' which is quantified in terms of measurements of the latter.  
Our results indicate that a deparametrized time evolution can be constructed based on coarse-graining 
localized observations in a classical reparametrization-invariant system which is ergodic. 

In previous related work it has always been assumed that global 
features of a trajectory are accessible to the observer, which makes it possible, in principle, to 
express the evolution of an arbitrarily selected degree of freedom relationally in terms of others 
\cite{MRT99,Montesinos00}. Thereby the Hamiltonian and possibly additional constraints have been eliminated 
in favour of Rovelli's ``evolving constants of motion'' \cite{Rovelli90}. 

This has been motivated by attempts to quantize gravity, based on the classical theory 
of General Relativity with its complicated constraint structure \cite{Rovelli90,Lawrie96,Peres97,Ohkuwa99}. 
In distinction, we pointed out in Ref.\,\cite{ES02} that the emergent discrete time in our approach naturally 
leads to a ``stroboscopic'' quantization of the system. Further arguments for a 
deterministically induced quantization have recently been proposed, for example, 
in Refs.\,\cite{Wetterich02,tHooft01}.   
  
The possibility of a fundamentally discrete time (and possibly other discrete 
coordinates) has been explored before, ranging from an early realization of Lorentz 
symmetry in such a case \cite{Snyder} to detailed explorations of its consequences 
and consistency in classical mechanics and quantum field theory \cite{TDL83,JN97}. 
However, no detailed models giving rise to such discreteness have been proposed and 
quantization has always been performed in an additional step, as usual.  
  
Here we summarize the distinctive new aspects of our simple model as follows: 
1. {\it Insisting on quasi-local measurements in describing the evolution, which respect reparametrization invariance 
of the system, the physical time necessarily becomes discrete; its construction is related to a Poincar\'e section 
of the ergodic dynamics.} 2. {\it Due to the inaccessability of globally complete 
information on trajectories, the  
evolution of remaining degrees of freedom appears as in a quantum mechanical model 
when described in relation to the discrete time.}    

We consider the (5+1)-dimensional model of a ``timeless'' relativistic particle 
(rest mass $m$) with the action:  
\begin{equation}\label{S}
S=\int\mbox{d}s\;L
\;\;, \end{equation}
where the Lagrangian is defined by: 
\begin{equation}\label{L}
L\equiv -\frac{1}{2}(\lambda^{-1}\dot x_\mu\dot x^\mu +\lambda m^2)
\;\;. \end{equation}
Here $\lambda$ stands for an arbitrary ``lapse'' function of the evolution parameter $s$, 
$\dot x^\mu\equiv\mbox{d}x^\mu /\mbox{d}s\,(\mu =0,1,\dots,5)$, and the metric is 
$g_{\mu\nu}\equiv\mbox{diag}(1,-1,\dots,-1)$. Units are such that $\hbar =c=1$.    

With this form of the Lagrangian, instead of the frequently encountered 
$L\mbox{d}s\propto (g_{\mu\nu}\mbox{d}x^\mu\mbox{d}x^\nu )^{1/2}$ 
which emphasizes the geometric (path length) character of the action, the presence of a constraint is immediately 
obvious, since there is no $s$-derivative of $\lambda$. 

Two spatial coordinates, $x^{4,5}$ in Eq.\,(\ref{L}), are toroidally compactified: 
\begin{eqnarray}\label{compact1}
x^{4,5}&\equiv&2\pi R[\phi^{4,5}] 
\;\;, \\ [1ex] \label{compact2}
[\phi ]&\equiv&\phi -n\;\;,\;\;\;\phi\in [n,n+1[
\;\;, \end{eqnarray} 
for any integer $n$, i.e. the angular variables are periodically continued; 
henceforth we set $R=1$, for convenience.
Alternatively, we can normalize the angular variables to the square $[0,1[\times [0,1[$, 
of which the opposite boundaries are identified, thus describing the surface of a torus with main radii equal to one.  

While full Poincar\'e invariance is broken, as in other currently investigated models 
with compactified higher dimensions, the usual one remains in fourdimensional Minkowski space together 
with discrete rotational invariance in the presently two extradimensions; also 
translational symmetry persists. Furthermore,  
the internal motion on the torus is ergodic with an uniform asymptotic density 
for almost all initial conditions, in particular if the ratio of the corresponding initial momenta 
is an irrational number \cite{Schuster}. 
    
Setting the variations of the action to zero, we obtain:  
\begin{eqnarray}\label{lapse}
\frac{\delta S}{\delta\lambda}&=&
\frac{1}{2}(\lambda^{-2}\dot x_\mu\dot x^\mu -m^2)=0
\;\;, \\ [1ex] \label{x}
\frac{\delta S}{\delta x_\mu}&=&
\frac{\mbox{d}}{\mbox{d}s}(\lambda^{-1}\dot x^\mu)=0
\;\;. \end{eqnarray}  
In terms of the canonical momenta, 
\begin{equation}\label{momentum} 
p_\mu\equiv\frac{\partial L}{\partial\dot x^\mu}=-\lambda^{-1}\dot x_\mu 
\;\;, \end{equation} 
the equations of motion (\ref{x}) become simply $\dot p^\mu =0$, while Eq.\,(\ref{lapse}) 
turns into the mass-shell constraint $p^2-m^2=0$. 

The equations of motion are solved by: 
\begin{equation}\label{xsolution}  
x^\mu (s)=x_i^\mu -p^\mu\int_0^s\mbox{d}s'\lambda (s')\equiv x_i^\mu +p^\mu\tau(s)
\;\;, \end{equation} 
where the conserved (initial) momentum $p^\mu$ is constrained to be on-shell and 
$x_i$ denotes the initial position. Here we 
also defined the fictitious proper time (function) $\tau$, which allows us to 
formally eliminate the lapse function $\lambda$ from Eqs.\,(\ref{lapse})-(\ref{x}), 
using $x^\mu (\tau )\equiv x^\mu (s)$ and  
$\dot x^\mu (s)=-\lambda (s)\partial_\tau x^\mu (\tau )$.    
  
In order to arrive at a physical space-time 
description of the motion, the proper time  
needs to be determined in terms of observables. In the simplest case, the result should be given by 
functions $x^{\mu\neq 0}(x^0)$, provided there is a physical clock measuring  
$x^0=x_i^0+p^0\tau$.  
  
Similarly as in the nonrelativistic example studied in Ref.\,\cite{ES02}, 
the lapse function introduces a gauge degree 
of freedom into the dynamics, which is related to the reparametrization of the evolution parameter $s$. 
In fact, the action, Eqs.\,(\ref{S})-(\ref{L}), is invariant under the set of gauge 
transformations: 
\begin{equation}\label{gauge} 
s\equiv f(s')\;\;,\;\;\; x^\mu (s)\equiv x'^\mu (s')\;\;,\;\;\; \lambda (s)\frac{\mbox{d}s}{\mbox{d}s'}\equiv\lambda'(s')
\;\;. \end{equation}   
It can be shown that the corresponding infinitesimal tranformations actually generate the evolution 
of the system (sometimes leading to the statement that there is no time in similar cases  
where dynamics is pure gauge). 
We do not discuss this further here, but refer to Ref.\,\cite{ES02} for a detailed 
and analogous discussion.   
  
Instead, with an evolution obviously taking place in such systems, 
we conclude from these remarks that the space-time description of motion requires 
a gauge invariant construction of a suitable time, replacing the fictitious proper time 
$\tau$. To this we add the important requirement that such construction should be based on 
quasi-local measurements, since global information (such as invariant path length)   
is generally not accessible to an observer in more realistic, typically nonlinear or higherdimensional theories.  

\section{``Timing'' through a window to extradimensions}   
Our construction of a physical time is based on the assumption that an observer 
in (3+1)-dimensional Minkowski 
space can perform measurements on full (5+1)-dimensional trajectories, 
however, only within a quasi-local window to the two extradimensions. 
In particular, the observer records the incidents 
(``units of change'') when the full trajectory hits an idealized detector which 
covers a small convex area element on the torus  
(compactified coordinates $x^{4,5}$).\footnote{One 
could invoke the distinction between brane and bulk matter 
as in string theory inspired higherdimensional cosmology, in order to construct more realistic 
models involving local interactions.} 

Thus, our aim is to construct time as an emergent   
quantity related to the increasing number of incidents measured by 
the reparametrization invariant incident number: 
\begin{equation}\label{incidentnumber}
I\equiv\int_{s_i}^{s_f}\mbox{d}s'\lambda (s')D(x^4(s'),x^5(s'))
\;\;, \end{equation} 
where $x^{4,5}$ describe the trajectory of the particle in the extradimensions, 
the integral is taken over the interval which corresponds to a 
given invariant path $x^\mu_i\rightarrow x^\mu_f$,  
and the function $D$ represents the detector features. 
Operationally it is not necessary to know the invariant path, in 
order to count the incidents.  

In the 
following examples we choose for $D$ the characteristic function 
of a small square of area $d^2$,  
$D(x^4,x^5)\equiv C_d(x^4)C_d(x^5)$, with $C_d(x)\equiv\Theta (x)(1-\Theta (x-d))$, 
which could be placed arbitrarily. Our 
results will not depend on the detailed shape of this idealized detector, if  
it is sufficiently small. More precisely, an incident is recorded only when, for 
example,  
the trajectory either leaves or enters the detector, or according to some other 
analogous restriction which could be incorporated into the 
definition of $D$. Furthermore, in order not to undo 
records, we have to restrict the lapse function $\lambda$ to be (strictly) 
positive, which also avoids trajectories which trace themselves backwards (or stall) 
\cite{Lawrie96}. 
The records correspond to  
a uniquely ordered series of events in Minkowski space, which are counted, and only 
their increasing total number is recorded, which is the Lorentz invariant 
incident number.    
  
Considering particularly the free motion on the torus, solution (\ref{xsolution}) yields:  
\begin{equation}\label{torusmotion}
[\vec\phi (\tau )]=[\vec\phi_0+\vec\pi\tau ] 
\;\;, \end{equation} 
where $\vec\phi$ is the vector formed of the angles $\phi^{4,5}$, and correspondingly 
$\vec\pi$, with 
$\pi^{4,5}\equiv p^{4,5}/2\pi R$; the quantities in Eq.\,(\ref{torusmotion}) are  
periodically continued, as before, see Eqs.\,(\ref{compact1})-(\ref{compact2}). 
Without loss of generality we choose  
$\vec\phi_0=0$ and $\pi^{5}>\pi^{4}>0$, 
and place the detector next to the origin with 
edges aligned to the positive coordinate axes for simplicity. 

Since here we are not interested in what happens between the incidents, we reduce 
the description of the internal motion to coupled maps.  
For proper time intervals 
$\Delta\tau$ with $\pi^4\cdot\Delta\tau=1$, the $\phi^4$-motion is replaced by the map 
$m\longrightarrow m+1$, where $m$ is a nonnegative integer, while: 
\begin{equation}\label{phi5}  
[\phi^5]=[Pm]
\;\;, \end{equation} 
with $P\equiv\pi^5/\pi^4>1$. 
Then, also the detector response counting incidents can be 
represented as a map:  
\begin{equation}\label{I}
I(m+1)=I(m)+\Theta (\delta -[\phi^5])+\Theta ([\phi^5]-(1-P\delta ))
\;\;, \end{equation}
with $I(0)\equiv 1$, and where $\delta\equiv d/2\pi R$ corresponds to the detector edge length $d$, 
assumed to be sufficiently small, $P\delta\ll 1$. The two $\Theta$-function contributions 
account for the two different edges through which the trajectory can enter the 
detector in the present configuration. 
  
The nonlinear two-parameter map (\ref{I}) has surpring universal features, some of 
which we will explore shortly. First of all, following the reparametrization invariant 
construction up to this point, we finally identify the {\it physical time} $T$ in terms  
of the incident number $I$ from Eq.\,(\ref{I}): 
\begin{equation}\label{time} 
T\equiv \frac{I}{\delta (\pi^4+\pi^5)}  
\;\;. \end{equation}   
A statistical argument for the scaling factor $\delta^{-1}(\pi^4+\pi^5)^{-1}$, based on  
ergodicity, has been 
given in Ref.\,\cite{ES02}, which applies here similarly. 

We show in Figure\,1  
how the physical time T typically is correlated with the fictitious proper time $\tau$. The  
proper time is extracted as those $m$-values when incidents happen: 
$\tau =m+1\Longleftrightarrow I(m+1)-I(m)=1$, corresponding to a particularly simple     
specification of the detector response. For a sufficiently small detector other such  
specifications yield the same results as described here. This is achieved by always rounding the  
extracted proper time values to integers, 
involving negligible errors of order $\delta,\delta /P\ll 1$. 

\begin{figure}[h]
\centerline{\hbox{
\psfig{figure=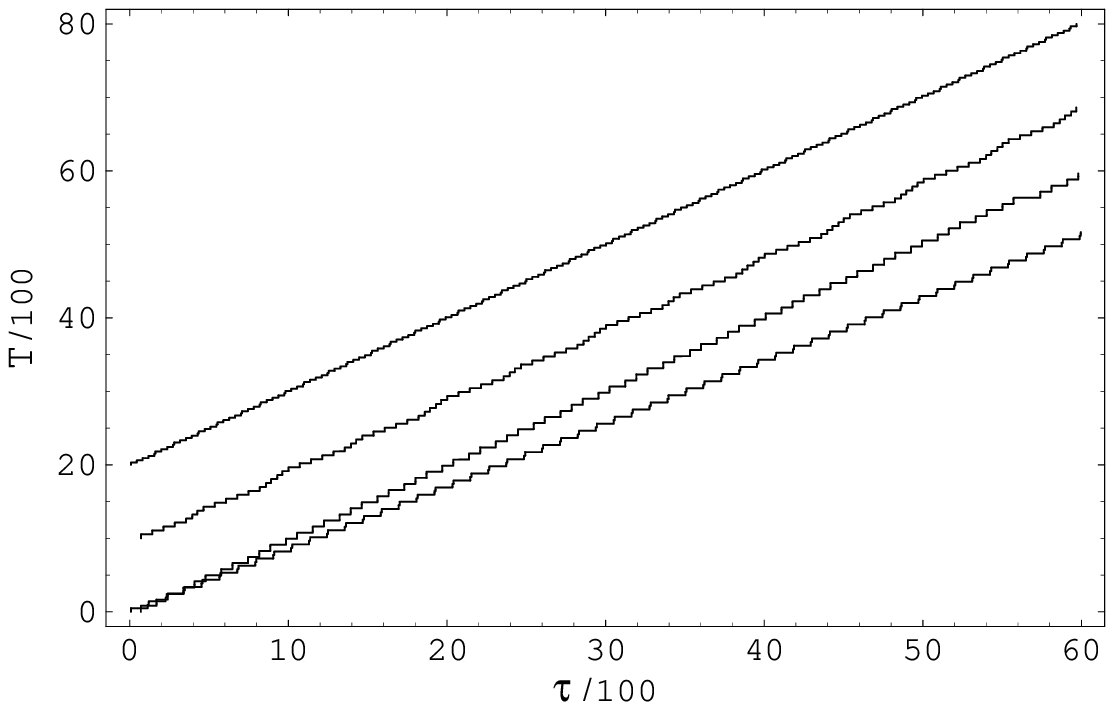,height=7.5cm,angle=0,clip=t}}}
\caption{The physical time $T$ as a function of proper time $\tau$ with detector 
parameter $\delta =.005$ and ratio of initial internal momenta $P=\sqrt{31}\;(\mbox{top}), e, \sqrt{2}, \pi\;(\mbox{bottom})$ (see main text); 
upper two curves displaced upwards by +10 and +20 units, respectively, for better visibility.}
\vskip 0.4cm      
\centerline{\hbox{
\psfig{figure=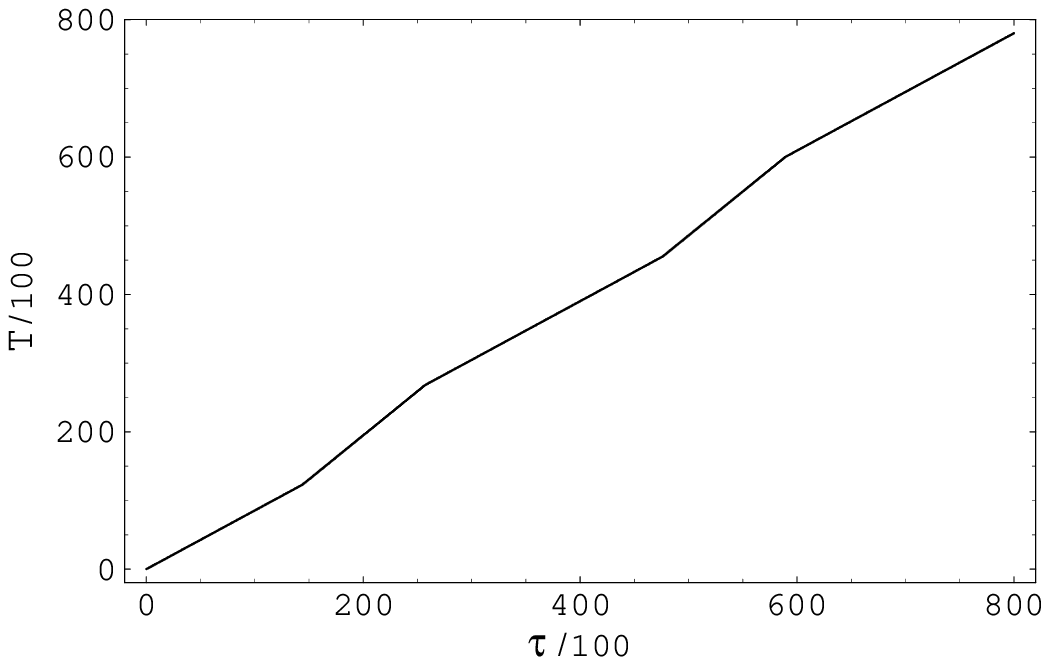,height=7.5cm,angle=0,clip=t}}}
\caption{Same as Figure\,1, with $P=\pi$, for much longer times.}
\end{figure}      
 
We find that the time $T$ does not run smoothly. This is due to the   
coarse-grained description of the internal motion: as if we were reading an analog clock  
under a stroboscopic light. In our construction, it is caused by the reduction of the full  
motion to a map (Poincar\'e section), corresponding to the recording of the physical incidents  
by the quasi-local detector.  

Furthermore, already after a short while, i.e. at low 
incident numbers, the constructed time approximates well the 
proper time $\tau$ on average. 
The fluctuations on top of the observed linear dependence 
result in the {\it discreteness} of the constructed time. 
  
Before we turn to the consequences of discreteness of the physical time, we illustrate in more 
detail its properties following from the present model. Note that the lowest curve in 
Figure\,1 appears to have a decidedly different slope from the others, apparently 
indicating a different scaling behavior than in Eq.\,(\ref{time}). However, as shown in 
Figure\,2, it is recovered after a sufficiently long time. 

In order to proceed, we explicitly solve  
the map, Eq.\,(\ref{I}),  
for the incident number as a function of the proper time (extracted as above), i.e. $m$.   
We obtain:   
\begin{equation}\label{incidents} 
I(m)=\sum_{k=0}^{m}{\cal C}(k)\equiv\sum_{k=0}^{m}\left ( 
\delta (P+1)+[Pk-\delta ]-[P(k+\delta )]\right )  
\;\;. \end{equation}  
Here we introduced the characteristic ``detector function'' ${\cal C}$ which equals 1 when there is an  
incident and is 0 otherwise. In this way, the detector response as a function of proper time is 
conveniently coded in a bit string.      

\begin{figure}[h]
\centerline{\hbox{
\psfig{figure=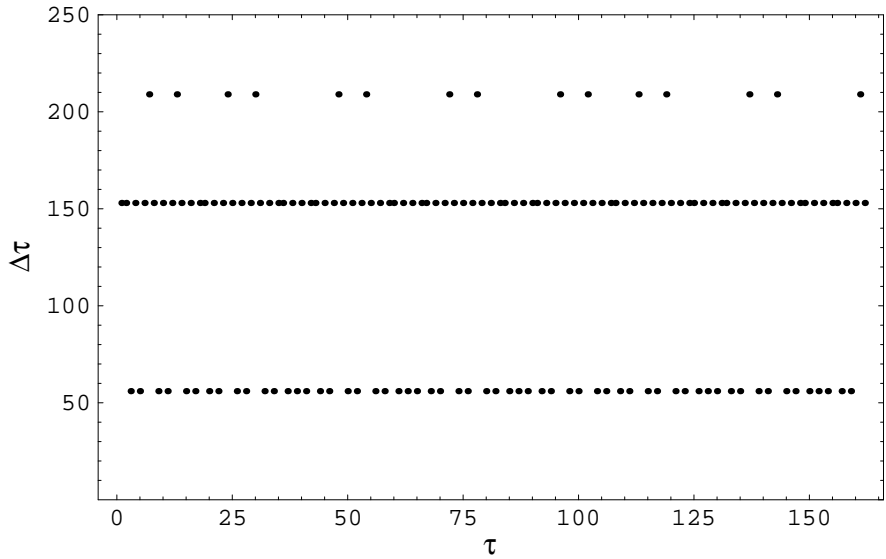,height=7.5cm,angle=0,clip=t}}}
\caption{The ``dead time'' $\Delta\tau$  
between consecutive incidents as a function of proper time $\tau$ with $\delta =.003$ and $P=\sqrt{3}$. Note the 
different frequency of appearance of the three values and $\Delta\tau_1\;(\mbox{bottom})+\Delta\tau_2\;(\mbox{middle})=\Delta\tau_3\;(\mbox{top})$.}
\end{figure}

Shown in a typical example in Figure\,3 is the integer time difference $\Delta\tau$ 
(``dead time'') between consecutive incidents as a function of proper time $\tau$. There are {\it always} three different values, $\Delta\tau_1<\Delta\tau_2<\Delta\tau_3$, which depend sensitively on the parameters $d$ and $P$. In particular, decreasing $d$ changes the relative frequency of occurrence of the $\Delta\tau_i$. Thus, at a critical value of $d$ either $\Delta\tau_1$ or $\Delta\tau_2$ disappears and simultaneously a new $\Delta\tau'$ appears. This annihilation respectively creation at critical parameter values continues with decreasing $d$. However, we find that the relation $\Delta\tau_1+\Delta\tau_2=\Delta\tau_3$ holds {\it always} among the three ordered values. This means that the characteristic detector function ${\cal C}$ generates a {\it two-frequency quasiperiodic process} \cite{Schuster}, showing interesting criticality features.    
  
The consecutive dead times define a discrete function, $\alpha (k)\equiv\Delta\tau (k),\;k=1,2,\dots\;$, which 
we employ in the autocorrelation function: 
\begin{equation}\label{autocorr} 
A(\Delta\tau )\equiv\frac{1}{M}\sum_{k=1}^M\left (
\alpha (k+\Delta\tau )\alpha (k)-\bar\alpha^2\right ) 
\;\;, \end{equation} 
where the average is $\bar\alpha\equiv\frac{1}{M}\sum_{k=1}^M\alpha (k)$, and ideally the limit $M\longrightarrow\infty$ should be taken.   

\begin{figure}[h]
\centerline{\hbox{
\psfig{figure=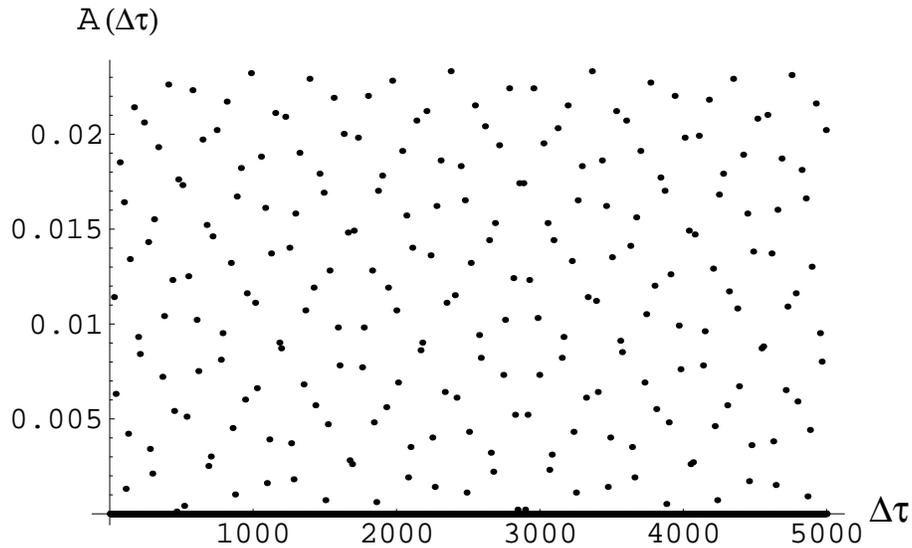,height=7.5cm,angle=0,clip=t}}}
\caption{The autocorrelation function $A(\Delta\tau)$, Eq.\,(\ref{autocorr}), with $\delta =.01$ and 
$P=\sqrt{2}$. Note the rich structure, which depends sensitively on the parameters, see Figures\,5,\,6.}
\end{figure}       

\begin{figure}[h]
\centerline{\hbox{
\psfig{figure=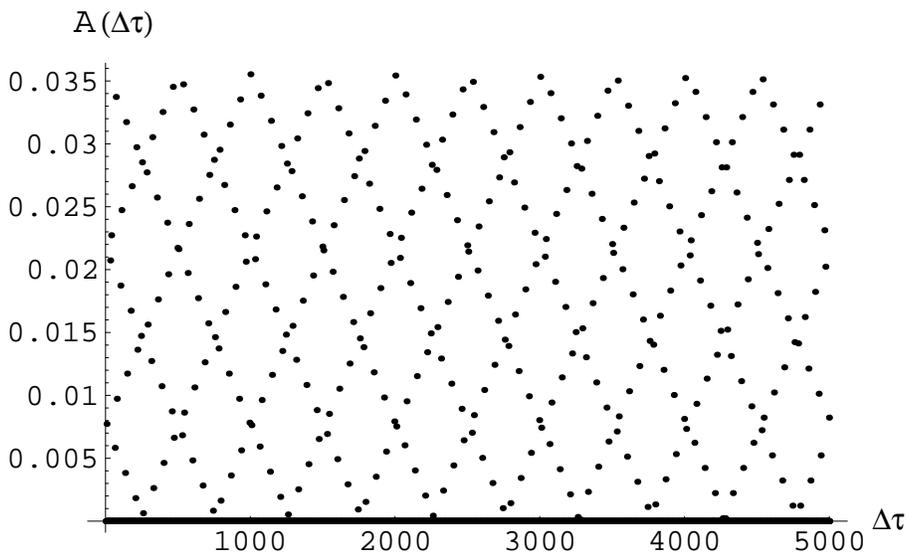,height=7.5cm,angle=0,clip=t}}}
\caption{Same as Figure\,4, with $P=e$.}
\end{figure} 
     
\begin{figure}[h]
\centerline{\hbox{
\psfig{figure=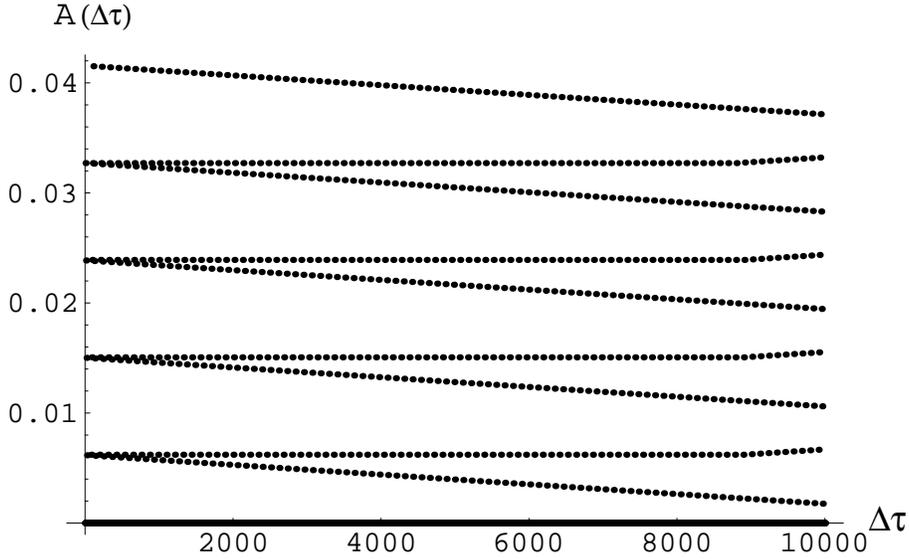,height=7.5cm,angle=0,clip=t}}}
\caption{Same as Figure\,4, with $P=\pi$.}
\end{figure} 
     
In Figures\,4--6 we show examples of $A$. The surprising feature here is that the autocorrelation function 
is neither decaying\footnote{In Figure\,6 there is a linear decay. However, in this case, still longer simulations 
might lead to a recovery of the autocorrelation function; cf. also Figure\,2.} 
nor constant or simply oscillating, as expected for either chaotic or regular processes, respectively \cite{Schuster}. 
Instead scatter plots of this discrete function reveil a rich structure which, again, depends sensitively on 
the parameters of the detector function  ${\cal C}$.   
Related features are found in the Fourier transform of the autocorrelation function, i.e., in the power spectrum of 
the process, which we do not study further here. 
    
\section{Emergent ``stroboscopic'' quantization} 
Following the construction of the discrete physical time through incident counting, 
we consider here the implications for the description of the motion of the 
external degrees of freedom in Minkowski space.   
  
After suitable translations and rotations, denoting the relevant coordinates by $x$ and $t$ 
and the associated momenta by $p$ and $p^0$, respectively, we have from Eq.\,(\ref{xsolution}) that   
$x=p\tau$ and $t=p^0\tau$, as usual. Since $x=(p/p^0)t$, it is sufficient to concentrate 
on $x(\tau )$, where the proper time $\tau$ is now to be understood as a function of the physical time $T$, Eq.\,(\ref{time}).  

It is obvious that a stochastic component arises in this description 
of the Minkowski space motion. It is caused by the fluctuations on top of the approximately linear behavior 
of $\tau (T)$, cf. Figure\,1, for example. 
 
We remark that any reasonable construction based on suitable exclusive observables 
should yield a total number of recorded incidents which increases in unit steps. 
These are the ``ticks of the clock'' which result in 
the time sequence $\{T_n,\;n\in\mathbf{N}\}$. Then, as we have seen in our model, the proper times $\tau (T_n)$ vary irregularly. This is due to the fact that monitoring a continuous ergodic internal motion with a localized detector generally produces a discrete quasi-periodic or more strongly irregular process, similarly as a Poincar\'e section. 

In particular, the discreteness of the time sequence generated in a generic nonintegrable system \cite{Schuster} is plausible for two reasons. First, by realistically restricting the observables to be localized near to the observer, we loose information about the full trajectories in coordinate space. Second, by insisting on gauge invariant observables, we loose the possibility to make use of additional momentum space observables (except for a limited number of constants of motion), or higher proper time derivatives, in order to reconstruct the full trajectories and a smooth evolution parameter based on them. 

Anyhow, for given initial conditions, the incident counts as well as the corresponding 
values of the Minkowski space coordinates present discrete physical data which are deterministically determined. The latter evolve stochastically. Then, the question arises, can we still give an invariant meaning to and predict the future of the free external motion, instead of letting the system evolve and measuring? - In general, the answer is `No', since the value of $\tau$ at the future instants $T_n$ is not predictable. 

However, following Ref.\,\cite{ES02}, we introduce the notion of a well-behaved time, which is sufficient to maintain some predictability indeed. 

Let a {\it well-behaved time} $T$ be such that the onedimensional motion of a free particle, co-ordinated in a reparametrization-invariant way by the sequence $\{ x(T_n)\}$, is of limited variation: 
\begin{eqnarray}
&\;&\{T_n\}\;\;\mbox{is well-behaved} 
\nonumber \\ [1ex] 
&\;&\Longleftrightarrow\;\;\exists\;\bar\tau ,\Delta\tau:\;
\bar\tau n-\Delta\tau\leq\tau (T_n)\leq\bar\tau n+\Delta\tau\;,\;\;\forall n\in{\mathbf N}
\nonumber \\ [1ex] \label{perfect2}
&\;&\Longleftrightarrow\;\;\exists\;\;\bar x,\Delta x:\;
\bar xn-\Delta x+x_i\leq x(T_n)\leq\bar xn+\Delta x+x_i\;,\;\;\forall n\in{\mathbf N}
\;\;, \end{eqnarray} 
with $\bar x\equiv p\bar\tau,\;\Delta x\equiv p\Delta\tau $, and where $x_i,p$ denote the initial data. This is the 
next-best we can expect from a reasonable time, relaxing the Newtonian concept. It can   
be exemplified in our present model. However, it has to be stressed that there is no hope for 
continuity or periodicity in the  
context of reparametrization invariant systems and local observables. 

Assuming a well-behaved time, we see that the sequence of points $\{ x(T_n)\}$ can be mapped into a regular lattice of possibly overlapping cells of size $2\Delta x$ and spacing $\bar x$. ``As the clock ticks'', i.e. $n\rightarrow n+1$, the particle moves from one cell to the next. On the average this takes a proper time $\bar\tau$ and physical time $\bar T$. Presently, we have $\bar T=\delta^{-1}(\pi_1^{(0)}+\pi_2^{(0)})^{-1}$, showing a dependence on the initial momenta and the scale of localization of the observed incidents. 

Thus, for a well-behaved time sequence, we have mappings from ``clock ticks'' to space and time intervals in Minkowski space, maintaining a fixed relation among them, however. As before, we identify space intervals, containing the respective $x(T_n)$, with {\it primordial states} \cite{ES02}: 
\begin{equation}\label{states}
|n)\equiv [\bar xn-\Delta x+x_i,\bar xn+\Delta x+x_i] 
\;\;. \end{equation} 
In order to handle the subtle limit of an infinite system, we impose reflecting boundary conditions and distinguish 
left- and right-moving states. Finally, then, the evolution is described by the deterministic rules of a cellular 
automaton: 
\begin{eqnarray}\label{R1}
n&\longrightarrow&n+1\;,\;\;n\in{\mathbf N}  
\;\;, \\ [1ex]
\label{R2}
|n)&\longrightarrow&|n+1)\;,\;\;-N+1\leq n\leq N-1
\;\;, \\ [1ex]
\label{R3}
|N)&\longrightarrow&|-N+1)
\;\;, \end{eqnarray}
with $2N$ states in all, $|-N+1),\;\dots\;,|N)$; states with a negative (positive) label correspond to left- (right-)moving states, according to negative (positive) particle momentum. Reaching the states $|0)$ or $|N)$, the particle changes its direction of motion. 

The rules (\ref{R1})--(\ref{R3}) can altogether be represented by a unitary evolution operator: 
\begin{equation}\label{U}
U(\delta T =\bar T)\equiv e^{-iH\bar T}=e^{-i\pi (N+1/2)/N}\cdot
\left (\begin{array}{ccccc}
0&\;&\;&\;&1 \\
1&0&\;&\;&\; \\
\;&1&0&\;&\; \\
\;&\;&\ddots&\ddots&\; \\
\;&\;&\;&1&0 
\end{array}\right )
\;\;, \end{equation} 
which acts on the $2N$-dimensional vector composed of the primordial states; the overall phase factor is introduced for later convenience. Here $\bar T$ is the natural scale for the Hamiltonian $H$. 

The formal aspects of following analysis are identical to the case of Ref.\,\cite{ES02}, applying 't\,Hooft's 
method \cite{tHooft01}, and we only give a summary of the results here. 
 
The evolution operator is diagonal with respect to the discrete Fourier transforms of the states $|n)$, which leads us to the basis functions: 
\begin{equation}\label{kBasis} 
\langle k|n)\equiv f_k(n)\equiv (2N)^{-1/2}\exp (\frac{i\pi kn}{N})   
\;\;, \end{equation} 
presenting a complete orthonormal basis. Next, we replace the states $|k\rangle$ by states $|m\rangle$, with 
$m\equiv -k+1/2$ and $-s\leq m\leq s$, 
where $2s+1\equiv 2N$. 

Then, recalling the algebra of $SU(2)$ generators, and with $S_z|m\rangle =m|m\rangle$ in particular, we obtain the Hamiltonian: 
\begin{equation}\label{Hdiag} 
H=\frac{2\pi}{(2s+1)\bar T}(S_z+s+\frac{1}{2}) 
\;\;, \end{equation} 
which is diagonal with respect to $|m\rangle$-states of the half-integer representations determined by $s$. The phase factor of Eq.\,(\ref{U}) contributes the additional terms which assure a positive definite (bounded) spectrum, reminding us of the harmonic oscillator. Furthermore, we have: 
\begin{equation}\label{Hsq} 
H=\frac{2\pi}{(2s+1)^2\bar T}(S_x^{\;2}+S_y^{\;2}+\frac{1}{4})+\frac{\bar T}{2\pi}H^2 
\;\;, \end{equation} 
using $S^2\equiv S_x^{\;2}+S_y^{\;2}+S_z^{\;2}=s(s+1)$. 

With $S_\pm\equiv S_x\pm iS_y$, we introduce coordinate and conjugate momentum operators: 
\begin{equation}\label{qp}
q\equiv\frac{1}{2}(aS_-+a^\ast S_+)\;\;,\;\;\; 
p\equiv\frac{1}{2}(bS_-+b^\ast S_+)
\;\;, \end{equation} 
where $a$ and $b$ are complex coefficients, chosen to satisfy $\Im (a^\ast b)\equiv -2(2s+1)^{-1}$. 
This already determines the basic commutator:  
\begin{equation}\label{commutator} 
[q,p]=i(1-\frac{\bar T}{\pi}H)
\;\;, \end{equation} 
since $[S_+,S_-]=2S_z$, and by Eq.\,(\ref{Hdiag}). 

In the infinite system limit ($s\rightarrow\infty$), a reasonably simple Hamiltonian follows, if we finally set: 
\begin{equation}\label{ab} 
a\equiv i\sqrt{\frac{\bar T}{\pi}}\;\;,\;\;\;b\equiv\frac{2}{2s+1}\sqrt{\frac{\pi}{\bar T}}
\;\;. \end{equation} 
Then, defining $\omega\equiv 2\pi /(2s+1)\bar T$, the previous Eq.\,(\ref{Hsq}) becomes: 
\begin{equation}\label{Hsq1} 
H=\frac{1}{2}p^2+\frac{1}{2}\omega^2q^2+\frac{\bar T}{2\pi}(\frac{1}{4}\omega^2+H^2)
\;\;, \end{equation} 
i.e. a nonlinearly modified harmonic oscillator Hamiltonian. 

Following our construction of the reparametrization-invariant physical time, presently $\bar T$ is finite in the infinite system limit. Thus, for $s\rightarrow\infty$, we have $\omega\rightarrow 0$ and obtain: 
\begin{equation}\label{Hpsq}
H=\frac{\pi}{\bar T}\left (1-(1-\frac{\bar T}{\pi}p^2)^{1/2}\right ) 
\;\;, \end{equation} 
which has the low-energy limit $H\approx p^2/2$. On the other side, the energy is bounded from above by $\pi /\bar T$, since we have $(\bar T/\pi )p^2=4(2s+1)^{-2}S_x^{\;2}$ ($\leq 1$, for $s\rightarrow\infty$, when diagonalized). 

Interestingly, towards the upper bound the violation of the basic quantum mechanical commutation relation, Eq.\,(\ref{commutator}), becomes maximal, i.e., there $q$ and $p$ commute like classical observables. Generally, however, 
the right-hand side interpolates between the usual quantum mechanical term and the classical limit. 

Finally, we give the matrix elements of the operators $q$ and $p$ with respect to   
primordial states. In the limit $s\rightarrow\infty$, the results are particularly 
simple:\footnote{Here we correct the obvious misprinting in two corresponding equations of Ref.\,\cite{ES02}.} 
\begin{eqnarray}\label{qmatrix} 
(n'|q|n)&=&\frac{1}{2}\sqrt{\pi\bar T}\frac{J_1(\pi (n'-n))}{n'-n}\frac{n'+n}{2}
\stackrel{n'\rightarrow n}{=}\frac{1}{4}\pi^{3/2}\bar T^{1/2}n 
\;\;, \\ [2ex]  
\label{pmatrix} 
(n'|p|n)&=&\frac{1}{2}\sqrt{\frac{\pi}{\bar T}}\frac{J_1(\pi (n'-n))}{n'-n}
\stackrel{n'\rightarrow n}{=}\frac{1}{4}\pi^{3/2}\bar T^{-1/2} 
\;\;, \end{eqnarray} 
where $J_1$ denotes an ordinary Bessel function of the first kind. Thus, neither the position nor the momentum operator is diagonal in the basis of the primordial states. However, a simple relation holds for the diagonal matrix elements, 
$(n|q|n)=n\bar T(n|p|n)$, in accordance with naive expectation. 

To summarize, we find here an emergent quantum model based on a deterministic classical evolution, similar as 
in Ref.\,\cite{tHooft01}. In particular, the relativistic particle model here presents  
a straightforward modification of the nonrelativistic model studied 
before \cite{ES02}. Since the relation between the external space and time coordinates 
is strictly maintained, $x\propto t$, the corresponding doubling of the stroboscopically 
quantized degrees of freedom is only apparent and does not lead to an intrinsically 
different quantum model: it is sufficient to introduce a onedimensional set of 
primordial states, $\{|n)\}$, since evolution of the time and space coordinates exactly follow  
each other.  

\section{Conclusions} 
Our ongoing construction of reparametrization-invariant time is based on the observation that ``time passes'' when there is an observable change, which is localized with the observer. More precisely, necessary are incidents, i.e. observable unit changes, which are recorded, and from which invariant quantities characterizing the change of the evolving system can be derived. 

Presently, we employ a window to compactified extradimensions in which the particle moves in addition 
to the relativistic motion in Minkowski space. More specifically, we consider a quasi-local 
detector which responds to the particle trajectory passing by in a simple Yes/No fashion. Counting  
such incidents, we construct an invariant measure of time. 

A basic ingredient is the assumption of ergodicity, such that the system explores dynamically the whole allowed energy surface in phase space. Generally, then there will be sufficiently frequent usable incidents next to the observer. They reflect properties of the dynamics with respect to (subsets of) Poincar\'e sections. Roughly, the  
passing time corresponds to the observable change there. We find that the particle's proper time 
is linearly related to the physical time, however, subject to stochastic fluctuations. 

It is most important to realize that the reparametrization-invariant time    
based on quasi-local observables naturally induces the stochastic  
features in the behavior of the external relativistic particle motion. Due to  
quasi-periodicity (or, generally, more strongly irregular features) of the emerging discrete time,  
the remaining predictable aspects can then most simply be described by a unitary  
discrete-time quantum mechanical evolution.  

This stroboscopic quantization seems to arise under more general circumstances, if one  
insists to construct (possibly by statistical means) an invariant time from localized observables.              
   
It will be most interesting to generalize the present study to fully diffeomorphism invariant systems, such as string models. Specifically deterministic classical systems with intrinsically many, even if freely moving degrees of freedom 
promise to give rise to emergent quantum models which differ from the ones found so far \cite{ES02,tHooft01}. 


\subsection*{Acknowledgements} 
I thank Christof Wetterich for discussions and all members of the Institut f\"ur Theoretische Physik 
for their kind hospitality. The support in parts by CNPq (Brasilia) 690138/02-4 and DAAD (Bonn) A/03/17806 
is gratefully acknowledged.  




\end{document}